\documentclass[aps,prev,twocolumn,preprintnumbers,floatfix,nofootinbib]{revtex4-1}
\pdfoutput=1
\usepackage[english]{babel}
\usepackage{graphicx,color}
\usepackage{bm}
\usepackage{times}
\usepackage{slashed}
\usepackage{multirow}
\usepackage[usenames,dvipsnames,svgnames,table]{xcolor}
\usepackage{slashed}
\usepackage{amsmath}
\usepackage{amsthm}
\usepackage{subfigure}
\usepackage{hyperref}
\definecolor{bluencs}{rgb}{0.0, 0.53, 0.74}
\definecolor{darkcyan}{rgb}{0.0, 0.55, 0.55}
\definecolor{hanblue}{rgb}{0.27, 0.42, 0.81}
\definecolor{blue2}{RGB}{53, 56, 170}
\usepackage{makecell}
\newcommand{\orcid}[1]{\hspace{1mm}\href{https://orcid.org/#1}{\includegraphics[height=0.3cm,keepaspectratio]{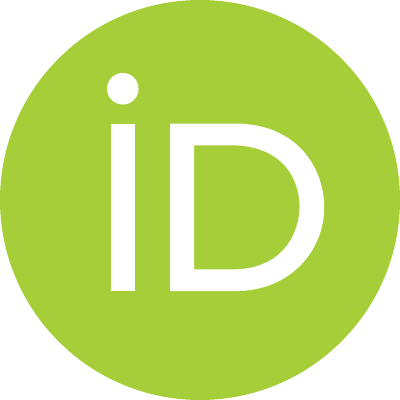}}}

\usepackage{booktabs}
\usepackage{makecell}      
\usepackage{siunitx}        
\usepackage{adjustbox}       
\usepackage{graphicx}
\hypersetup{
	linktocpage=true,
	setpagesize=true,
	urlcolor=darkcyan,
	citecolor=darkcyan,
	linkcolor=darkcyan, 
	menucolor=cyan,
	colorlinks=true, 
	filecolor=darkcyan,
	citebordercolor=darkcyan,
	pdftitle={Exploring Potential Higgs Resonances at 650 GeV and 95 GeV in the 2HDM Type III},
	pdfsubject={Latex},
	pdfauthor={Benbrik, Boukidi, Moretti, Kahime, Rahili, Taki},
	pdfkeywords={BSM},
	unicode=true
}
\usepackage[capitalize]{cleveref}

\newcommand{\be}{\begin{equation}}
\newcommand{\ee}{\end{equation}}
\newcommand{\bea}{\begin{eqnarray}}
\newcommand{\eea}{\end{eqnarray}}

\usepackage{xfrac}
 
\usepackage{float}
\usepackage{bbding,pifont}
\definecolor{brilliantrose}{rgb}{1.0, 0.33, 0.64}
\definecolor{lawngreen}{rgb}{0.49, 0.99, 0.0}
\definecolor{magenta}{rgb}{1.0, 0.0, 1.0}
\makeatletter
\let\old@float\@float
\def\@float{\let\centering\relax\old@float}
\makeatother
\begin{document}
	
	\title{Exploring Potential Higgs Resonances at 650 GeV and 95 GeV in the 2HDM Type III}

	\author{R. Benbrik$^{1}$\orcid{0000-0002-5159-0325}}
	\email{r.benbrik@uca.ac.ma}	
	\author{M. Boukidi$^{1,2}$\orcid{0000-0001-9961-8772}}
	\email{mohammed.boukidi@ced.uca.ma}
	\author{K. Kahime\orcid{0000-0001-8013-3521}$^{3}$}
	\email{Kahimek@gmail.com}
	\author{S. Moretti$^{4,5}$\orcid{0000-0002-8601-7246}}
	\email{s.moretti@soton.ac.uk}\email{stefano.moretti@physics.uu.se}
		\author{L. Rahili$^{6}$\orcid{0000-0002-1164-1095}}
	\email{rahililarbi@gmail.com}
		\author{B. Taki$^{6}$\orcid{0009-0009-2642-1288}}
	\email{taki.bassim@edu.uiz.ac.ma}

	\affiliation{$^1$Polydisciplinary Faculty, Laboratory of Fundamental and Applied Physics, Cadi Ayyad University, Sidi Bouzid, B.P. 4162, Safi, Morocco.\\	$^2$Institute of Nuclear Physics, Polish Academy of Sciences, ul. Radzikowskiego 152, Cracow, 31-342, Poland.\\	$^3$Laboratoire Interdisciplinaire de Recherche en Environnement, Management, Energie et Tourisme (LIREMET), ESTE, Cadi Ayyad University, B.P. 383, Essaouira, Morocco.\\	$^4$School of Physics and Astronomy, University of Southampton,\\ Southampton, SO17 1BJ, United Kingdom.\\$^5$Department of Physics and Astronomy, Uppsala University, Box 516, SE-751 20 Uppsala, Sweden.\\$^6$Laboratory of Theoretical and High Energy Physics (LPTHE), Faculty of Sciences, Ibnou Zohr University, B.P 8106, Agadir, Morocco.}

\begin{abstract}
Recent searches by the CMS collaboration at the Large Hadron Collider (LHC) in the `diphoton plus $b\bar{b}$ final state'  have revealed an excess near 650 GeV, which might indicate the presence of a (broad) heavy resonance decaying into a Standard Model (SM) Higgs boson and an additional particle. At the same time, both ATLAS and CMS as well as all experiments at the 
Large Electron-Position (LEP) collider have reported  excesses consistent with a  light Higgs boson channel around 95 GeV, suggesting a potential connection between these two signals. In this study, we investigate these anomalies within the context of the CP-conserving 2-Higgs-Doublet Model (2HDM) Type~III. In our proposed scenario, a heavy CP-odd Higgs boson $A$ with a mass near 650 GeV decays into a SM-like Higgs boson $H$ and a $Z$ boson, with the $H$ boson subsequently decaying into diphotons and the $Z$ boson decaying into $b\bar{b}$ pairs. To explain the excess around 95 GeV, we independently select the mass of the light CP-even Higgs boson of the model $h$ to be around this value. This configuration allows for a consistent explanation of both the high- and low-mass excesses observed in the experiments. Through a detailed analysis, we demonstrate that these two signals can be fitted simultaneously at the 2.5$\sigma$ significance level, offering a promising solution to the observed anomalies and potentially hinting at new physics Beyond the Standard Model (BSM).

\end{abstract}
%%%%%%%%%%%%%%%%%%%%%%%%%%%%%%%%%%%%%%%%%%
\maketitle
	\section{Introduction}
\label{sec:intro}

The discovery of the 125 GeV Higgs boson at the Large Hadron Collider (LHC)~\cite{ATLAS:2012yve,CMS:2012qbp} has firmly confirmed the Electro-Weak (EW) Symmetry Breaking (EWSB) mechanism described by the Standard Model (SM) and validated the central role of the Higgs field in giving mass to fundamental particles. This milestone confirmed the last missing piece of the SM, yet it simultaneously opened the door to further investigations. In fact, while the SM provides a comprehensive framework for particle interactions, it is also known to have several limitations, particularly in explaining phenomena like Dark Matter (DM), neutrino masses and the hierarchy problem. These shortcomings motivate the exploration of well-motivated extensions to the SM.

Among the most prominent theoretical frameworks are Supersymmetric models, like the Minimal Supersymmetric Standard Model (MSSM)~(see Ref.~\cite{Moretti:2019ulc} for a review), as a high energy theory, and 2-Higgs-Doublet Models (2HDMs), as low energy framework~\cite{Gunion:1992hs,Branco:2011iw}, both of which predict the existence of additional (pseudo)scalar particles beyond the SM Higgs boson. The search for these additional states, which could play a crucial role in addressing the above open questions, remains one of the central objectives of both ongoing and future experimental programmes at the LHC and other experimental facilities.

A recent CMS search in the`$\gamma\gamma$ (diphoton) plus $b\bar{b}$' final state has reported an intriguing excess at approximately 650 GeV, with a local significance of 3.8$\sigma$~\cite{CMS:2023boe}. This excess has been interpreted as evidence of a heavy resonance decaying into a SM-like Higgs boson and an additional state, potentially (but not necessarily) a new particle within an extended Higgs sector. The measured production rate, expressed as the gluon-gluon fusion  into the 650 GeV state cross section times the relevant Branching Ratios (${\cal BR}$s), is approximately 0.35 fb, which suggests that this signal deviates from what is expected in the SM  in this decay channel.

Simultaneously, several independent analyses have reported potential signals of a light Higgs-like state near 95 GeV, further fueling interest in the exploration of extended Higgs sectors. Notably, the LEP collaborations observed a mild excess in the $e^+e^-\to Z^*\to Zb\bar{b}$ channel in the mass range of 95–100 GeV~\cite{ALEPH:2006tnd}. At the LHC, CMS initially reported an excess of approximately 2$\sigma$ in the diphoton channel near 97 GeV during Run 1~\cite{CMS:2015ocq}, which was later reinforced by Run 2 data, yielding a local significance of 2.9$\sigma$, specifically, at $m_{\gamma\gamma}=95.4$ GeV\cite{CMS:2018cyk,CMS:2024yhz}. In parallel, an independent search by ATLAS in the diphoton channel observed a similar excess with a local significance of 1.7$\sigma$~\cite{ATLAS:2024bjr,ATLAS:2023jzc}. Moreover, CMS also observed a 2.6$\sigma$ excess near 95 GeV in the di-tau channel~\cite{CMS:2022goy}, although recent constraints on Higgs production in association with top-quark pairs or a $Z$ boson decaying into $\tau^+\tau^-$~\cite{CMS:2024ulc} make this result somewhat less definitive (so that we will ignore it in our forthcoming numerical analysis). Given these constraints, our analysis focuses primarily on the diphoton and diphoton plus $b\bar{b}$ signatures, where excesses in both the high- and low-mass regions have been consistently reported.

The simultaneous observation of excesses in distinct mass regions, 650 GeV and 95 GeV, has spurred interest in the possibility of unified theoretical models that can explain both mass excesses within a coherent framework. Several models have been proposed to explain the presence of additional (pseudo)scalar states, including the Next-MSSM (NMSSM)~\cite{Ellwanger:2023zjc}, the N2HDM-$U(1)_H$~\cite{Banik:2023ecr} and various other Beyond SM (BSM) scenarios~\cite{Cao:2016uwt,Heinemeyer:2021msz,Biekotter:2021qbc,Biekotter:2019kde,Cao:2019ofo,Biekotter:2022abc,Iguro:2022dok,Li:2022etb,Cline:2019okt,Biekotter:2021ovi,Crivellin:2017upt,Cacciapaglia:2016tlr,Abdelalim:2020xfk,Biekotter:2022jyr,Biekotter:2023jld,Azevedo:2023zkg,Biekotter:2023oen,Cao:2024axg,Wang:2024bkg,Li:2023kbf,Dev:2023kzu,Borah:2023hqw,Cao:2023gkc,Aguilar-Saavedra:2023tql,Ashanujjaman:2023etj,Dutta:2023cig,Ellwanger:2024txc,Diaz:2024yfu,Ellwanger:2024vvs,Ayazi:2024fmn,Coloretti:2023wng,Bhattacharya:2023lmu,Ahriche:2023hho,Ahriche:2023wkj,Benbrik:2022azi,Benbrik:2022tlg,Belyaev:2023xnv,Janot:2024lep,Gao:2024ljl,Benbrik:2024ptw,Li:2025tkm,Hmissou:2025uep,Gao:2024qag,Dutta:2025nmy,Abbas:2025ser,Xu:2025vmy,Arhrib:2025pxy,Coutinho:2024zyp,Abbas:2024jut,Baek:2024cco,Banik:2024ugs,Mondal:2024obd,Dong:2024ipo,Robens:2024wbw,BrahimAit-Ouazghour:2024img,Khanna:2024bah,Janot:2024ryq,YaserAyazi:2024hpj,Ogreid:2024qfw,Du:2025eop,Lian:2024smg}. Among these extensions, the 2HDM remains particularly attractive due to its minimality and its ability to naturally accommodate additional(pseudo)scalar particles. The Type~III variant of the 2HDM, where both Higgs doublets couple to all fermions in a CP-conserving manner, has been shown to effectively explain the observed excesses by implementing an appropriate Yukawa texture. This model provides a natural explanation for the 95 GeV signal in the form of a light CP-even Higgs boson~\cite{Benbrik:2022azi,Benbrik:2022tlg,Belyaev:2023xnv} (or even  a superposition of this and a CP-odd state~\cite{Benbrik:2024ptw}) 
while remaining consistent with the experimental constraints on other properties of the Higgs sector.

In this study, we focus on a CP-conserving 2HDM Type~III scenario, where a heavy CP-odd Higgs boson, $A$, with a mass near 650 GeV decays into a SM-like Higgs boson, $H$, and a $Z$ boson in the decay channel $A \to HZ$. The SM-like Higgs boson, $H$, decays into diphotons whereas the $Z$ boson decays into a $b\bar{b}$ pair, a signature which could explain the observed excess in the high-mass region. Simultaneously, we introduce a light CP-even Higgs boson, $h$, with a mass around 95 GeV, to account for the low-mass diphoton excess. Our analysis shows that this framework can simultaneously accommodate both observed excesses, yielding a combined significance of approximately 2.5$\sigma$ when considering all relevant experimental constraints. This result provides a potential solution to the observed anomalies, suggesting that both excesses may be consistent with a unified extended Higgs sector.

The remainder of this paper is organised as follows. In Section 2, we detail the theoretical framework of the CP-conserving 2HDM Type III, including the scalar potential and Yukawa interactions. Section 3 reviews the experimental excesses in the 650 and 95 GeV mass regions and summarises the key Higgs production and decay channels. In Section 4, we present the results of our numerical analysis, discussing the fit to the excesses and the constraints from various experiments. Finally, Section 5 concludes with a summary and outlook on future directions for research in this area.

%%%%%%%%%%%%%%%%%%%%%%%%%%%%%%%%%%%%%%%%%%%%%%%%%%%%%%%%%%%%%%%%%%%%%%%%%%%%%%%%%%%%%%%%%%%%%%%%  
%%%%%%%%%%%%%%%%%%%%%%%%%%%%%%%%%%%%%%% Model framework section %%%%%%%%%%%%%%%%%%%%%%%%%%%%%%%%

\section{The 2HDM Type III}
\label{sec:model}

We consider the CP-conserving 2HDM of Type III, which extends the Standard Model (SM) by introducing a second $SU(2)_L$ scalar doublet with hypercharge $Y = 1$. Electroweak symmetry breaking leads to five physical Higgs states: two CP-even ($h$, $H$), one CP-odd ($A$), and a charged pair ($H^\pm$). We focus on the inverted hierarchy where the heavier CP-even state $H$ corresponds to the observed 125~GeV Higgs boson.

{The scalar potential and Yukawa sector are described in Refs.~\cite{Belyaev:2023xnv,Benbrik:2024ptw}. In this framework, both doublets couple to all fermions, generating tree-level Flavour Changing Neutral Currents (FCNCs). To suppress such effects, we employ the Cheng-Sher ansatz~\cite{Cheng:1987rs,Diaz-Cruz:2004wsi,Hernandez-Sanchez:2012vxa}, wherein the Yukawa couplings scale as $\sqrt{m_i m_j}/v \times \chi^f_{ij}$, with $\chi^f_{ij}$ denoting real, dimensionless parameters.

For simplicity and to avoid Lepton Flavour Violation (LFV), we assume a flavour-diagonal texture by setting $\chi^f_{ij} = 0$ for $i \ne j$. This choice efficiently suppresses LFV processes such as $\mu \to e \gamma$, $\tau \to \mu \gamma$ and $\tau \to 3\mu$, while maintaining the mass-proportional structure of the ansatz and reducing the number of free parameters.}

Our analysis is carried out in the physical basis\footnote{We set $\lambda_6 = \lambda_7 = 0$ for simplicity.}, specified by the Higgs masses $m_h$, $m_A$, and $m_{H^\pm}$, the ratio of vacuum expectation values $t_\beta$, the CP-even mixing parameter $s_{\beta - \alpha}$, and the soft breaking parameter $m_{12}^2$. This is supplemented by the diagonal Yukawa parameters $\chi^f_{ii}$.

All parameter points considered are required to satisfy theoretical consistency conditions, including perturbative unitarity~\cite{uni1,uni2}, vacuum stability~\cite{Barroso:2013awa,sta}, and the perturbativity of quartic couplings~\cite{Branco:2011iw}. Experimental constraints include electroweak precision observables~\cite{Grimus:2007if}, LHC Higgs signal strength data evaluated with \texttt{HiggsSignals-v3}~\cite{Bechtle:2020pkv,Bechtle:2020uwn} via \texttt{HiggsTools}~\cite{Bahl:2022igd}, and exclusion bounds from LEP, Tevatron, and LHC Higgs searches using \texttt{HiggsBounds-v6}~\cite{Bechtle:2008jh,Bechtle:2011sb,Bechtle:2013wla,Bechtle:2015pma}. Additional flavor constraints from $B \to X_s \gamma$, $B_{s,d} \to \mu^+ \mu^-$, and $B^+ \to \tau^+ \nu_\tau$ are implemented using \texttt{SuperIso\_v4.1}~\cite{superIso}. The numerical analysis follows the methodology of Refs.~\cite{Belyaev:2023xnv,Benbrik:2024ptw}, ensuring full consistency with both theoretical and experimental bounds.

%%%%%%%%%%%%%%%%%%%%%%%%%%%%%%%%%%%%%%%%%%%%%%%%%%%%%%%%%%%%%%%%%%%%%%%%%%%%%%%%%%%%%%%%%%%%%%%%  
%%%%%%%%%%%%%%%%%%%%%%%%%%%%%%%%%%%%%%%  Analysis section %%%%%%%%%%%%%%%%%%%%%%%%%%%%%%%%%%%%%%

\section{Analysis of the Excesses in the $\gamma\gamma$ and $b\bar b$ Channels}
\label{sec:excess}

In this section, we examine the ability of the 2HDM Type III to simultaneously account for the observed excesses near 95~GeV and 650~GeV in the $\gamma\gamma$ and $b\bar b$ final states, both individually and in combination.

The combined excess can be interpreted as a heavy resonance decaying into a SM-like Higgs boson and another particle. Specifically, CMS reports a best-fit signal yield~\cite{CMS:2023boe} of
$$
\sigma_{\gamma\gamma b\bar b}^{\rm CMS} = 0.35^{+0.17}_{-0.13}~\text{fb}.
$$
In the 2HDM Type III, as intimated, $H_{\rm SM}$ is the heaviest CP-even Higgs state, $H$, whereas $X_{95}$ can represent either a light CP-even Higgs boson, $h$, or, in fact,  a $Z$ boson. Notice that the latter possibility, which we adopt here (so that the 95 GeV excesses can be
explained in terms of  the $h$ state), is plausible for two reasons. Firstly, the invariant mass resolution of the $b\bar b$ system is larger than 5 GeV. Secondly, the significance of the excess is maximised over the interval 70 GeV $<m_{b\bar b}<$ 1 TeV (or so)\footnote{This will require correcting the $Z$ decay rates to remove off-shell contributions of the gauge boson below the lower end of this mass interval, which we will account for in all forthcoming numerical studies (it is of order percent).}. Thus, in our framework, the $X_{650}$ is nothing but the CP-odd Higgs state, $A$.

{This reinterpretation is well motivated. Although the CMS analysis~\cite{CMS:2023boe} originally models the excess as a decay $X_{650} \to H_{125} Y_{95}$, with $Y$ being a spin-0 state, we consider instead the process $A \to Z H_{125}$. At $\sqrt s =m_A \simeq 650$~GeV, where $m_Z^2/s \ll 1$, the Goldstone  equivalence theorem ensures that the longitudinal polarisation of the $Z$ boson behaves like the corresponding neutral (pseudoscalar) Goldstone mode, with deviations suppressed by ${\cal O}(m_Z^2/s) \lesssim 2\%$.

Importantly, the CMS selection strategy relies on invariant mass reconstruction and general kinematic features of the $\gamma\gamma b\bar b$ system, without employing cuts that explicitly depend on spin (or polarisation). Therefore, any  difference 
in acceptance 
between (pseudo)scalar and longitudinal vector interpretations is expected to be negligible relative to the statistical uncertainty of the observed signal. As such, the reported CMS rate remains applicable to our reinterpretation.

Further support comes from the most recent CMS search for $A \to ZH$~\cite{CMS:2025bvl}, which sets 95\% Confidence Level (CL) upper bounds on the inclusive cross section at approximately 0.077 pb for gluon-fusion and 0.0705~pb for $b$-associated production when $m_A = 650$ GeV. As shown in Fig.\ref{fig5} (Appendix), the predicted cross sections corresponding to our scenario lie well below these experimental limits in both production channels. }

To describe the low-mass excess at 95~GeV, we define signal strength modifiers in the $b\bar b$ and $\gamma\gamma$ final states as

\begin{equation} \mu_{{b\bar{b}}} = \frac{\sigma_{\rm 2HDM}(e^+e^-\to Z\phi)}{\sigma_{\rm SM}(e^+e^-\to ZH^{95}_{\rm SM})} \times \frac{{\cal BR}_{\rm 2HDM}(\phi \to b\bar{b})}{{\cal BR}_{\rm SM}(H^{95}_{\rm SM} \to b\bar{b})}, \end{equation} where $\sigma_{\rm 2HDM}$ refers to the Higgs production rate and $\mathcal{BR}$ to the decay rate. Similarly, for the diphoton channel, the signal strength modifier is given by:

\begin{equation} \mu_{\gamma\gamma} = \frac{\sigma_{\rm 2HDM}(gg \to \phi)}{\sigma_{\rm SM}(gg \to H^{95}_{\rm SM})} \times \frac{{\cal BR}_{\rm 2HDM}(\phi \to \gamma\gamma)}{{\cal BR}_{\rm SM}(H^{95}_{\rm SM} \to \gamma\gamma)}. \end{equation}

Here, $\phi$ corresponds to the light CP-even Higgs boson, $h$, in our 2HDM Type III setup and $H^{95}_{\rm SM}$ is a fictitious CP-even Higgs state with SM couplings and a mass of 95 GeV.

The experimentally measured signal strengths for the 95 GeV excesses reported by the LHC (ATLAS and CMS combined~\cite{Biekotter:2023oen}) as well as LEP provide important data points for comparison with the 2HDM Type III predictions. The measured signal strengths for the $\gamma\gamma$ and $b\bar{b}$ channels are given by, respectively:

\begin{equation} \mu_{\gamma\gamma}^{\mathrm{exp}} = 0.24^{+0.09}_{-0.08}, \quad \mu_{b\bar{b}}^{\mathrm{exp}} = 0.117 \pm 0.057. \end{equation}

The uncertainties in these values are relatively large but they still suggest a possible presence of BSM signals, particularly when considered in conjunction with the excess observed at 650 GeV.
In this analysis, as mentioned, we consider the light CP-even Higgs boson, $h$, as a candidate to explain the observed excess in the diphoton and $b\bar{b}$ channels at 95 GeV. By comparing the $\gamma\gamma$ and $b\bar b$ signal strengths above to experimental data recasted in the same form as well as by fitting the event rate corresponding to the $b\bar b\gamma$ final state, we can then assess the viability of the 2HDM Type III as a unified explanation for all discussed excesses.

The following sections will delve deeper into the theoretical implications of these and other experimental results. Specifically, we will explore the parameter space of the 2HDM Type III in the presence of consistency requirements.

%%%%%%%%%%%%%%%%%%%%%%%%%%%%%%%%%%%%%%%%%%%%%%%%%%%%%%%%%%%%%%%%%%%%%%%%%%%%%%%%%%%%%%%%%%%%%%%%
%%%%%%%%%%%%%%%%%%%%%%%%%%%%%%%%%%  Numerical results section %%%%%%%%%%%%%%%%%%%%%%%%%%%%%%%%%%

\section{Explanation of the Excesses}
\label{sec:results}

We now present our numerical analysis to assess whether the 2HDM Type III can simultaneously account for the excesses observed at approximately 95\ GeV and 650\ GeV in the diphoton and $b\bar{b}$ channels as well as the $b\bar b\gamma\gamma$ final state, respectively.

Spectrum generation is performed using \texttt{2HDMC 1.8.0}~\cite{2HDMC}, which incorporates the theoretical constraints and EWPOs outlined above. The resulting parameter points are then subjected to experimental tests using \texttt{HiggsTools} and \texttt{SuperIso\_v4.1}.

In the scenario considered here, the heavier CP-even Higgs boson, $H$, is identified with the SM-like Higgs observed at the LHC with $m_{H}=125.09$ GeV. The lighter CP-even Higgs boson, $h$, with $m_{h}\in[94,97]$\ GeV, is responsible for the excesses observed in the diphoton and $b\bar{b}$ channels near 95\ GeV. Simultaneously, the heavy CP-odd Higgs boson, $A$, with $m_{A}\in[625,675]$\ GeV, is assumed to decay via $A\to HZ$, with $H\to\gamma\gamma$ and $Z\to b\bar{b}$, thereby generating the 650\ GeV excess.

As for the scan intervals of the other model parameters, these are listed in Tab.~\ref{tab:par-scan}.

\begin{table}[h!]
	\centering
		\renewcommand{\arraystretch}{1.2} 
	\begin{adjustbox}{max width=0.5\textwidth}
		\begin{tabular}{c|c|c|c|c|c|c|c}\hline\Xhline{0.85pt} \addlinespace[1pt]\Xhline{0.5pt}
			$m_h$ & $m_H$ & $m_A$ & $m_{H^\pm}$ & $s_{\beta-\alpha}$ & $t_{\beta}$ & $m_{12}^2$ & $\chi_{ij}^{f,\ell}$ \\\Xhline{1pt}
			$[94,\,97]$ & $125.09$ & $[625,\,675]$ & $[600,\,700]$ & $[0,\,0.5]$ & $[1,\,15]$ & $m_h^2s_{\beta-\alpha}c_{\beta-\alpha}$ & $[-3,\,3]$ \\\Xhline{0.5pt} \addlinespace[1pt]\Xhline{0.85pt}
		\end{tabular}
	\end{adjustbox}
	\caption{Scan ranges of the 2HDM Type III input parameters (masses in GeV).}
	\label{tab:par-scan}
\end{table}

\begin{figure}[htbp!]
	\centering
	\includegraphics[height=7.75cm,width=7.75cm]{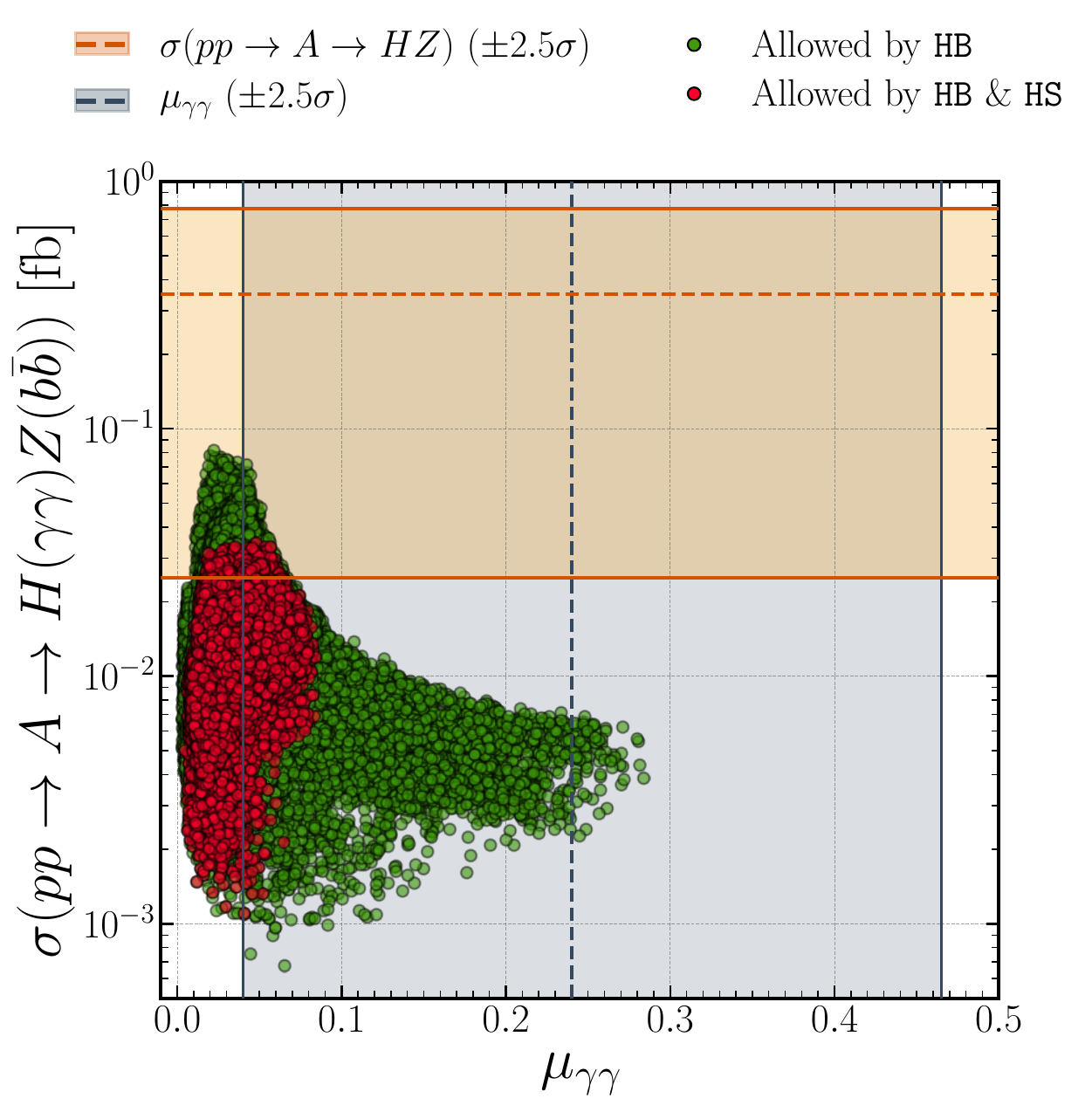}	
	\caption{Correlation between $\mu_{\gamma\gamma}$ and $\sigma(pp \to A \to H(\to \gamma\gamma) Z(\to b\bar{b}))$. The orange (gray) dashed line marks the central value of the cross section ($\mu_{\gamma\gamma}$), with the corresponding 2.5$\sigma$ uncertainty shown as a shaded band. Green points satisfy all theoretical and experimental constraints except {\tt HiggsSignals}, while red points additionally satisfy the {\tt HiggsSignals} test.}
	\label{fig1}
\end{figure}

\begin{figure}[htbp!]
	\centering
	\includegraphics[height=7.75cm,width=7.75cm]{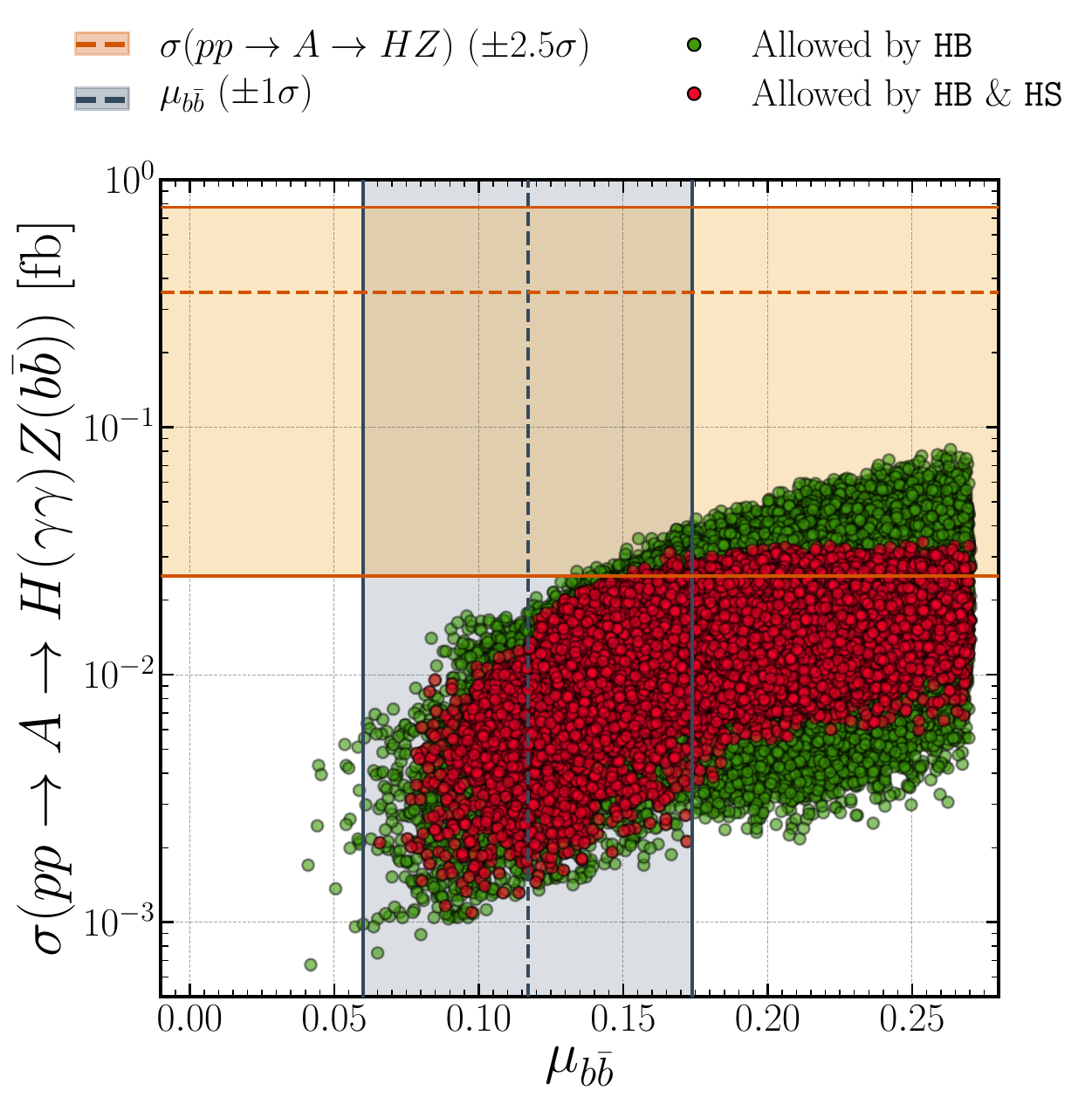}	
    \caption{Same as Fig.~\ref{fig1}, but for $\mu_{b\bar{b}}$. The orange dashed line and band represent the cross section and its 2.5$\sigma$ uncertainty, while the gray dashed line and band show the central value and 1$\sigma$ uncertainty on $\mu_{b\bar{b}}$.}
	\label{fig2}
\end{figure}

Figs.~\ref{fig1} and~\ref{fig2} display the predicted signal strengths $\mu_{\gamma\gamma}$ and $\mu_{b\bar{b}}$, respectively, as functions of the cross section $\sigma(pp\to A\to H(\to \gamma\gamma)Z(\to b\bar{b}))$. The dashed lines and bands indicate the experimental central values and 2.5$\sigma$ uncertainties. Red points represent parameter space points that satisfy all imposed constraints, including those from {\tt HiggsSignals}, while green points meet all constraints except for the latter requirements.

The results clearly indicate that the 2HDM Type III can simultaneously reproduce the observed excesses at 95~GeV and 650~GeV at the 2.5$\sigma$ level. In this framework, the decay chain $A \to HZ$, where $A$ is a heavy CP-odd Higgs and $H$ is a light CP-even state, provides a natural mechanism to explain both features: the light scalar$h$  accounts for the diphoton and $b\bar{b}$ excesses near 95~GeV, while the heavy state $A$ is responsible for the potential resonance around 650~GeV. The predicted signal strengths align well with CMS and ATLAS data, particularly in regions of parameter space where $\mu_{\gamma\gamma} \sim 0.04$--0.1 while $\mu_{b\bar{b}}$ remains consistent with LEP and LHC observations.

\begin{figure*}[htbp!]
	\centering
	\includegraphics[height=14.5cm,width=14.cm]{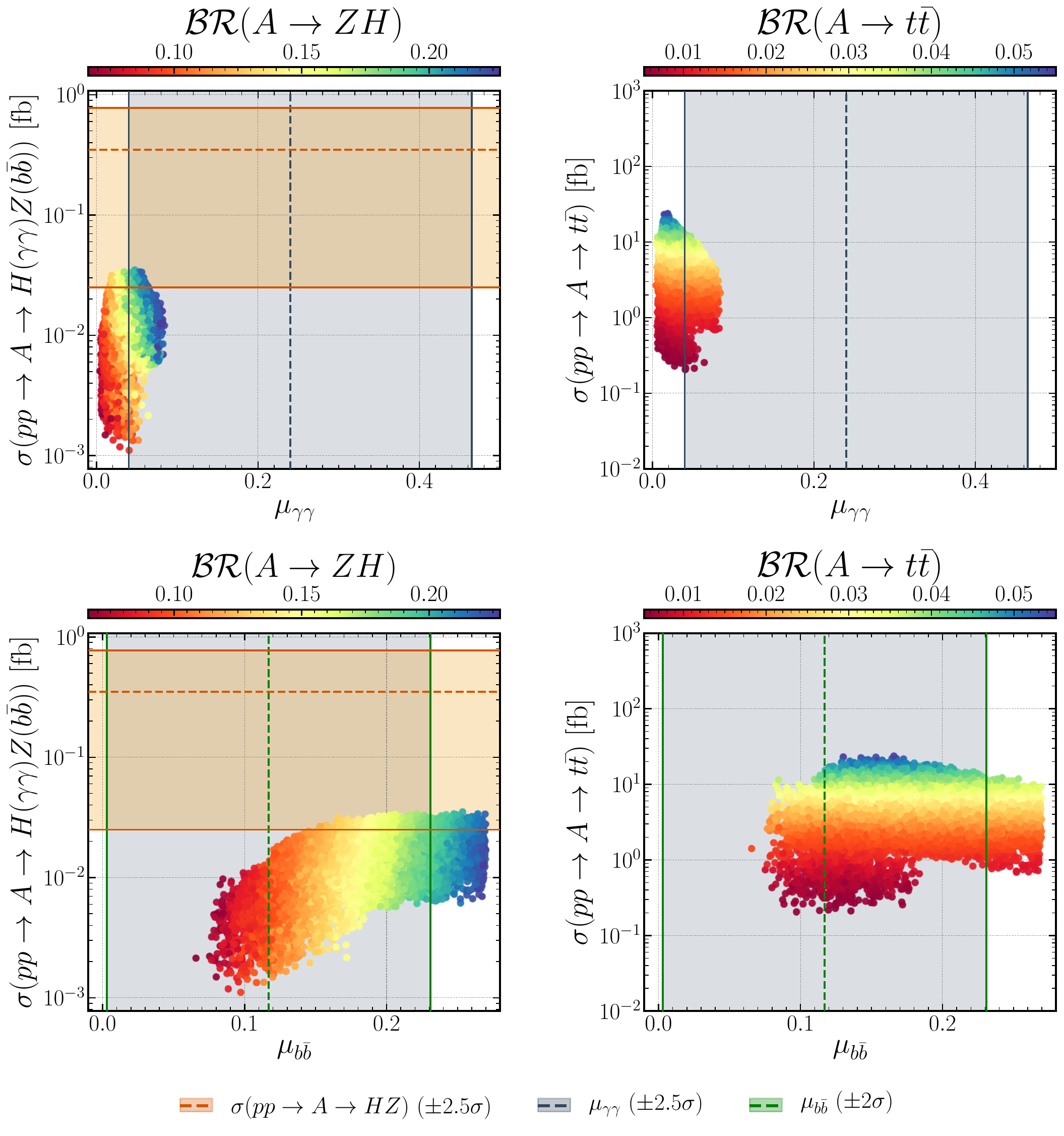}	
	\caption{Parameter space points satisfying all theoretical and experimental constraints. The upper (lower) panels display the correlations between $\mu_{\gamma\gamma}$ ($\mu_{b\bar{b}}$) and the cross sections $\sigma(pp \to A \to H(\to \gamma\gamma) Z(\to b\bar{b}))$ (left) and $\sigma(pp \to A \to t\bar{t})$ (right). The colour scale indicates $\mathcal{BR}(A \to t\bar{t})$ in the left panels and $\mathcal{BR}(A \to ZH)$ in the right panels.}\label{fig3}
\end{figure*}

To further explore the features of the viable 2HDM Type III  parameter space, Fig.~\ref{fig3} shows the ${\cal BR}$s of the heavy CP-odd Higgs boson $A$ into $Z H$ (left) and $t\bar{t}$ (right), shown as colour scales. The upper (lower) panels display correlations with $\mu_{\gamma\gamma}$ ($\mu_{b\bar{b}}$) in the $\sigma(pp \to A \to H(\to \gamma\gamma) Z(\to b\bar{b}))$ and $\sigma(pp \to A \to t\bar{t})$ planes. In the region compatible with both the 95~GeV and 650~GeV excesses (within 2.5$\sigma$), the $\mathcal{BR}(A \to Z H)$ stabilises around 20$\%$, supporting the interpretation that the $A \to Z H$ transition plays a dominant role in establishing the 650 GeV signal. Furthermore, the $A\to t\bar{t}$ decay mode, which becomes increasingly relevant at higher $m_A$ values, yields cross sections reaching up to $\approx12$~fb, still within the experimental limits from current LHC searches for heavy resonances decaying into $t\bar t$ final states.

These findings therefore illustrate a coherent picture in which a light scalar near 95~GeV and a heavier pseudoscalar around 650~GeV can naturally arise in the 2HDM Type III framework with extended Yukawa structures. Moreover, the ability of the model to accommodate both such excesses without violating theoretical and experimental constraints lends significant credibility to its viability. The strong correlations between production cross sections, ${\cal BR}$s, signal strengths (at 95 GeV) and event rates (at 650 GeV) reinforces the interpretative power of the proposed scenario and offer the chances to provide Benchmark Points (BPs) over its parameter space which can be tested at current and upcoming LHC runs.
\begin{table*}[t!]
	\centering
	\renewcommand{\arraystretch}{1.5} 
	\footnotesize
	\caption{Model parameters and predictions for the selected BPs. Masses are in GeV.}
	\begin{tabular}{ccccccc ccc ccc ccc cc cc}
		\Xhline{0.85pt}  \addlinespace[1pt]\Xhline{0.5pt}
		&{$m_h$} & {$m_H$} & {$m_A$} & {$m_{H^\pm}$} & {$s_{\beta-\alpha}$} & {$t_\beta$}
		& {$\chi_{11}^u$} & {$\chi_{22}^u$} & {$\chi_{33}^u$}
		& {$\chi_{11}^d$} & {$\chi_{22}^d$} & {$\chi_{33}^d$}
		& {$\chi_{11}^l$} & {$\chi_{22}^l$} & {$\chi_{33}^l$}
		& {$\mu_{\gamma\gamma}$} & {$\mu_{b\bar{b}}$}
		& {$\sigma_{pp \to A\to HZ}$~(fb)} & {$\sigma_{pp \to A\to t\bar t}$~(fb)} \\
		\Xhline{1pt}
		BP$_1$	&95.44 & 125.09 & 638.55 & 645.53 & 0.45 & 8.47 & 0.01 & 0.69 & -0.27 & 0.05 & 0.06 & 1.56 & -0.19 & -0.36 & 1.46 & 0.0410 & 0.2297 & 0.031 & 11.68 \\
		BP$_2$	&	95.34 & 125.09 & 649.96 & 650.29 & 0.44 & 8.19 & 0.18 & 0.70 & -0.28 & -0.13 & 0.16 & 1.56 & 0.25 & -0.40 & 1.43 & 0.0409 & 0.2295 & 0.028 & 11.24 \\
		\Xhline{0.5pt} \addlinespace[1pt]\Xhline{0.85pt}
	\end{tabular}\label{tab:BPS}
\end{table*}

\begin{figure*}[htbp!]
	\centering
	\includegraphics[height=12cm,width=14cm]{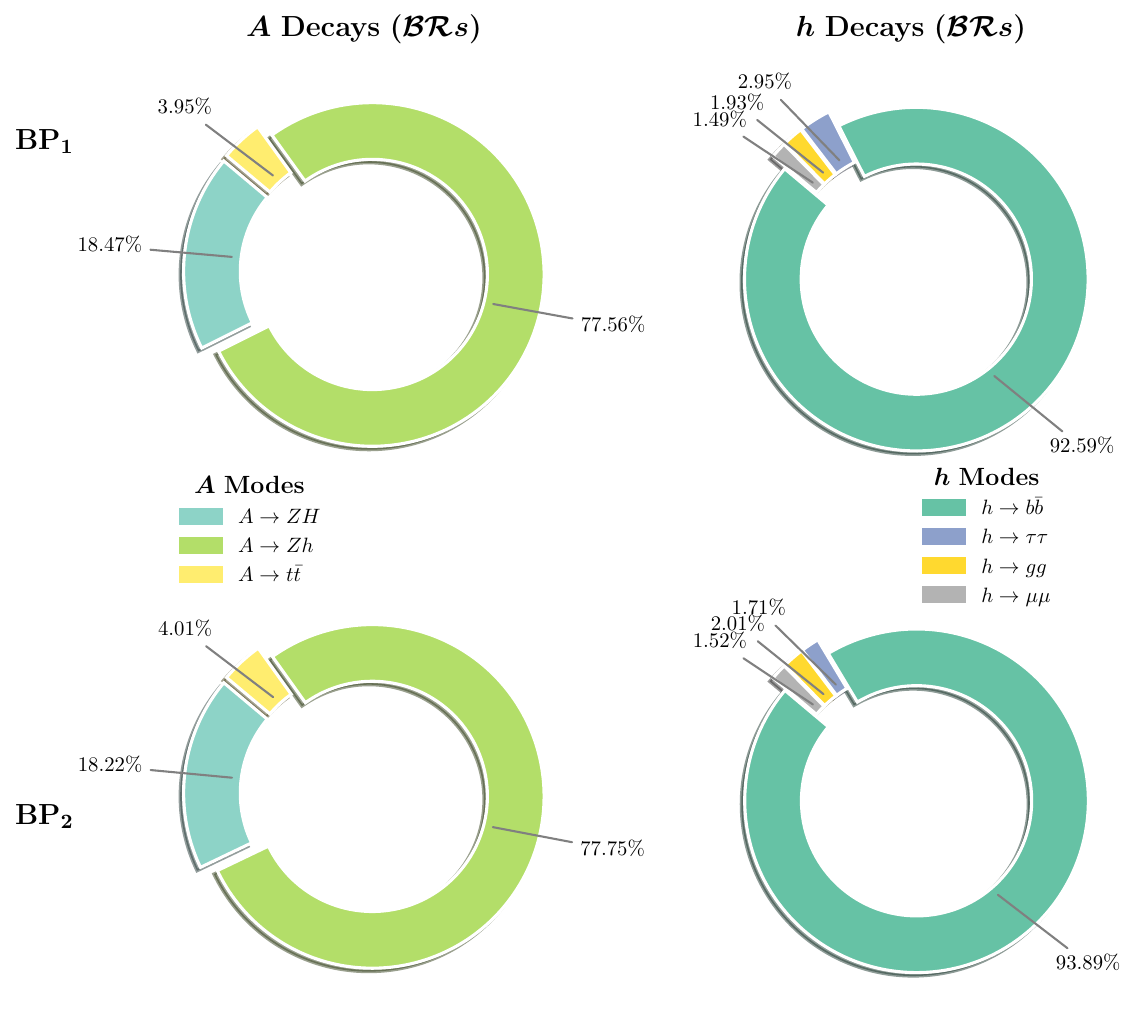}	
	\caption{$\mathcal{BR}$s of $A$ and $h$ decays for BP$_1$ and BP$_2$.}
	\label{fig4}
\end{figure*}

Therefore, Tab.~\ref{tab:BPS} presents two BPs\footnote{{The small values of $\chi_{11}^f$ are a consequence of scanning the parameter space for optimal agreement with the observed excesses. They are not enforced by constraints or tuning  and larger values remain allowed.}} that successfully reproduce the observed excesses near 95~GeV and 650~GeV. Their main features are as follows.

The signal strengths $\mu_{\gamma\gamma}$ and $\mu_{b\bar{b}}$ are fully consistent with experimental trends, while the predicted cross sections of 0.031~fb and 0.028~fb for the process $pp \to A \to H(\to \gamma\gamma)Z(\to b\bar{b})$ lie well within the sensitivity of current LHC analyses. Crucially, for both BPs, moderate values of $\tan\beta$ and a non-zero $s_{\beta-\alpha}$ allow sizable couplings of $h$ to SM particles. In particular, enhanced third-generation Yukawa couplings lead to significant contributions in the $h \to \gamma\gamma$ and $h \to b\bar{b}$ decay channels.

Finally, Fig.~\ref{fig4} shows the ${\cal BR}$s for the two BPs, highlighting $A \to Z h$ and $h \to b\bar{b}$ as the dominant decay modes that characterize these scenarios.
%%%%%%%%%%%%%%%%%%%%%%%%%%%%%%%%%%%%%%%%%%%%%%%%%%%%%%%%%%%%%%%%%%%%%%%%%%%%%%%%%%%%%%%%%%%%%%%%  
%%%%%%%%%%%%%%%%%%%%%%%%%%%%%%%%%%%%%%%  Conclusions %%%%%%%%%%%%%%%%%%%%%%%%%%%%%%%%%%%%%%%%%%%

\section{Conclusions}\label{con}

We have examined the potential of the CP-conserving 2HDM Type~III to simultaneously address two distinct experimental anomalies: the $\sim$95~GeV excesses observed in the diphoton and $b\bar{b}$ channels (the first from the LHC and the second from LEP)  and the $\sim$650~GeV excess in the $\gamma\gamma + b\bar{b}$ final state (also seen at the LHC, but most notably reported by the CMS Collaboration). In this framework, the light CP-even Higgs boson $h$ explains the low-mass excesses through its enhanced couplings to third-generation fermions and non-negligible interactions with gauge bosons, while the heavy CP-odd Higgs boson $A$ undergoes the decay chain $A \to HZ$ with $H \to \gamma\gamma$ and $Z \to b\bar{b}$, thereby accounting for the high-mass structure.
Our analysis incorporated the latest theoretical consistency conditions, including vacuum stability, perturbativity, and unitarity, as well as current experimental constraints from flavour physics, EWPOs and collider limits. Using state-of-the-art numerical tools, we have identified viable regions in the 2HDM Type III parameter space where the excesses are simultaneously accommodated at the 2.5$\sigma$ level.

Furthermore, we have provided BPs that exemplify the successful realisation of this BSM scenario, characterised by moderately large values of $\tan\beta$, sizeable Yukawa couplings to third-generation fermions as well as non-alignment, which allows for non-negligible $hVV$ couplings ($V=W^\pm, Z$). We have also explored the two competing decay channels of the heavy pseudoscalar $A$, particularly $A \to ZH$ and $A \to t\bar{t}$, highlighting regions of parameter space where the former is indeed contributing to the 650 GeV excess while the latter is such that direct searches for heavy resonances in $t\bar t$ final states are not constraining our viable regions. Needless to say, both    these modes can be probed simultaneously in future LHC analyses.

Overall, our results suggest that the 2HDM Type~III offers a consistent and predictive framework for addressing both low- and high-mass anomalies in current collider data. Future experimental efforts, including improved measurements in the diphoton and $b\bar{b}$ final states, separately as well as combined in the same final state, and dedicated searches for heavy scalar resonances decaying into $HZ$ and $t\bar{t}$, will be crucial in testing the viability of this scenario.

\section*{Acknowledgments}
\sloppy
RB is supported in part by the PIFI Grant No. 110200EZ52. MB acknowledges the support of Narodowe Centrum Nauki under OPUS grant no. 2023/49/B/ST2/03862. SM is supported in part through the NExT Institute and STFC CG ST/X000583/1. 

\appendix
\section*{Appendix}
\begin{figure*}[htbp!]
	\centering
	\includegraphics[height=7.75cm,width=15.cm]{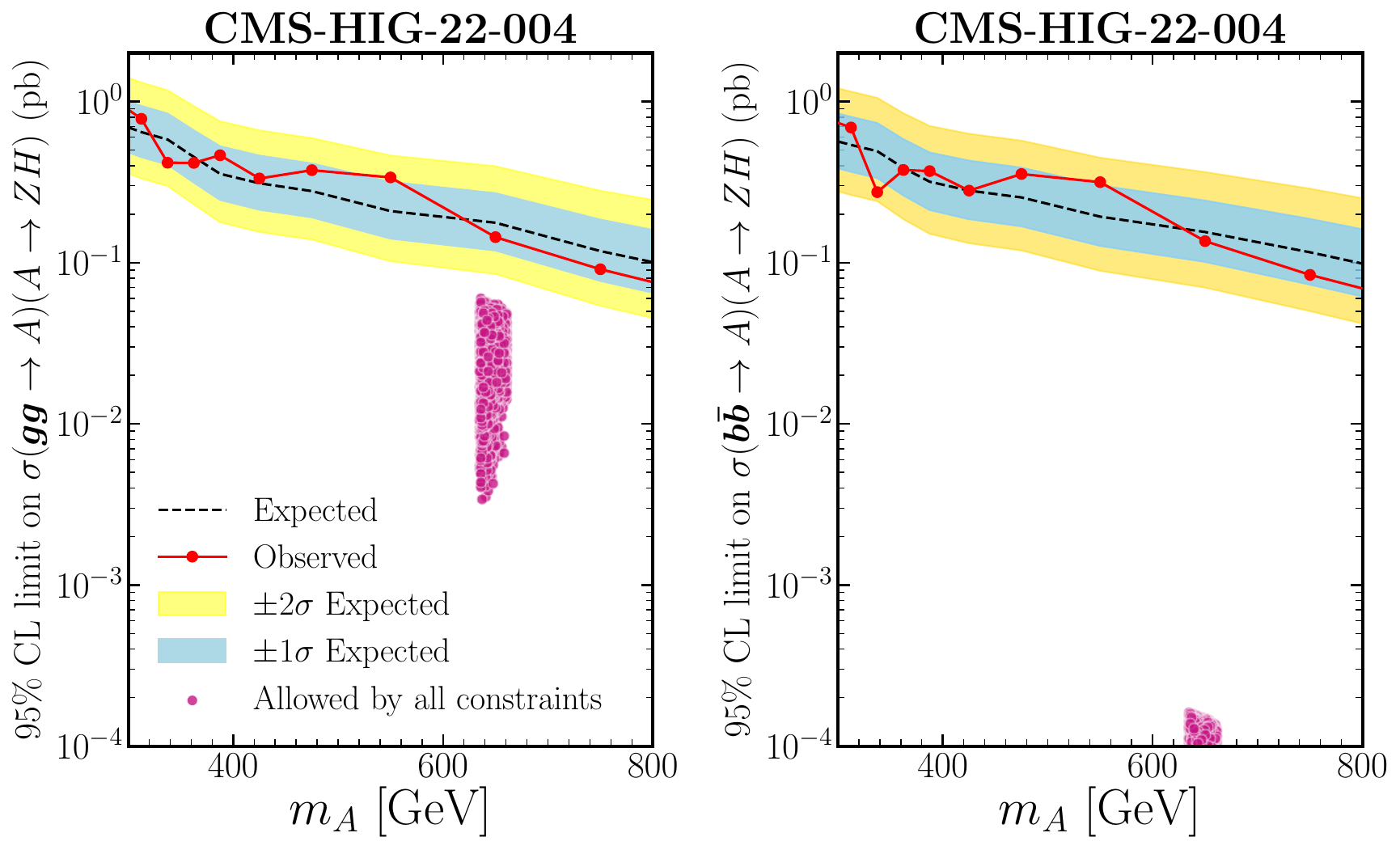}	
	\caption{Surviving parameter points for $\sigma \times \mathcal{B}(A \to Z H_{125})$, overlaid onto the observed 95\% CL CMS-HIG-22-004 exclusion limits~\cite{CMS:2025bvl}. Left: gluon-gluon fusion production ($gg \to A$). Right: $b$-associated production ($b\bar{b} \to A$). All points satisfy the full set of theoretical and experimental constraints discussed in the main text.
	}
	\label{fig5}
\end{figure*}

In Fig.~\ref{fig5}, we show the predicted values of $\sigma \times \mathcal{B}(A \to Z H_{125})$ for all viable points that satisfy the full set of theoretical and experimental constraints. These are superimposed onto the observed 95\% CL exclusion limits from the CMS search for $A \to Z h$~\cite{CMS:2025bvl}. The left panel corresponds to gluon-gluon fusion ($gg \to A$), and the right panel to $b$-associated production ($b\bar{b} \to A$).

As evident from the figure, all points lie significantly below the exclusion contours in both production channels, confirming the compatibility of our interpretation with current experimental constraints.

\bibliography{main} 
\bibliographystyle{apsrev4-2}
\end{document}